\documentstyle[12pt,epsf]{dubna98}

\begin{document}

\begin{center}
{\bf THE HYDROGEN LAMB SHIFT AND THE PROTON RADIUS\\}
\vspace*{1cm}
SAVELY G. KARSHENBOIM\footnote{E-mail: sek@mpq.mpg.de; sgk@onti.vniim.spb.su}\\
{\it D. I. Mendeleev Institute for Metrology, 198005 St. Petersburg,
Russia\\
and\\
Max-Planck-Institut f\"ur Quantenoptik, D-85748, Garching, Germany}
\end{center}
\newcommand{\beq}{\begin{equation}}
\newcommand{\eeq}{\end{equation}}
\newcommand{\eq}[1]{Eq. (\ref{#1})}
\vspace*{.5cm}
\begin{abstract}
{\small
The Lamb shift measurement and theory are now both a dynamically developing
field and we give a review of the current data. Critical comparison of theory and
experiment can be done using a value of the proton charge radius and we
pay attention to several results of its determination. 
}
\end{abstract}

\vspace*{1cm}
The talk is devoted to the Lamb shift in the {\em hydrogen\/} atom. The workshop is
for {\em exotic atoms\/} and first of all I would like to say
that to my mind the hydrogen {\em is\/} one of them. The possibility both to
calculate and to measure different energy intervals with extremely high
accuracy make hydrogen a quite exotic system. Next, one has to
remember that there is actually no special separate theory of the
{\em positronium\/} or {\em muonium\/} atom, 
investigations of which were discussed here in detail. 
The same 
expression can be of use in a
number of calculations for several atomic systems (see e. g. Refs. \cite{Fell,Pachu}. 
One more point which is quite similar to the {\em pionium\/} and 
{\em pionic hydrogen\/} 
is that we investigate an atomic system but
after all the result is important for particle physics, namely for the {\em proton
charge radius\/}. All of these reasons show that hydrogen atomic properties are
to be discussed among exotic ones. The workshop talk is based on a 
Max-Planck-Institut f\"ur Quantenoptik report \cite{review} 
and all {\em references\/} can be found there.

The hydrogen atom is one of the most important {\em QED\/} systems. In contrast
to muonium and positronium the nucleus is a proton and hence the energy
of the level is influenced by the proton structure, i. e. by the {\em strong\/}
interaction. The knowledge of the proton radius leads now to a limit
of the theoretical value for the Lamb shift. We discuss here the problem
of the radius determination, the most popular values of which are summarized
in Table 1. We give a critical review of all of these values, as well as of
the theory and the experiment on the Lamb shift.

\begin{table}[h]
\begin{center}
\begin{tabular}{||c|c|c||}
\hline\hline
&&\\[-1ex]
Value         & Reference  & Method \\  [1ex]
\hline
&&\\[-1ex]
0.809(11)~{\rm fm}  & Stanford, 1963 \protect{\cite{Hand}}
  & scattering \& empirical fitting \\ [1ex]
0.862(12)~{\rm fm} & Mainz, 1980     \protect{\cite{Simon}}
  &  scattering \& empirical fitting \\ [1ex]
0.64(8)~~~{\rm fm} & Draper {\sl et al.}, 1990  \protect{\cite{Draper}}
  & lattice QCD in chiral limit \\ [1ex]
0.88(3)~~~{\rm fm} & Leinweber, Cohen, 1993     \protect{\cite{Leinweber93}}
  & lattice QCD \& chiral perturbation  \\ [1ex]
0.847(9)~{\rm fm} &  Mainz, 1996   \protect{\cite{Mergell}}
 & dispersion relation fitting     \\ [1ex]
0.890(14)~{\rm fm} &  Garching, 1997  \protect{\cite{Udem}}
& hydrogen Lamb shift measurements \\[1ex]
\hline\hline
\end{tabular}
\end{center}\vskip-\lastskip
\caption{\em Proton charge radius \label{t1}}
\end{table}

Let us briefly describe all items of the Table 1. The elastic
electron-proton {\em scattering data\/} were obtained 
with small momentum transfer
and were extrapolated to zero momentum. The other way to extract the radius
from such data is based on a {\em dispersion relation approach\/} which allows
to involve data from other kinematic areas into fitting. The numerical 
calculation within the {\em chiral limit\/} of 
the {\em lattice QCD\/} can be corrected
due to the {\em chiral perturbative theory\/}. And indeed one of the ways 
to determine the radius is based on the hydrogen Lamb shift investigation.
First we mention that we expect that the uncertainties presented
in Table 1 accordingly to the original works \cite{Hand,Draper,Leinweber93} 
are significant underestimations and we will not consider those anymore
here. The details can be found in our review \cite{review}.

We start our consideration with the Lamb shift. Several results obtained
within different approaches are summarized in Table 2 (see Fig. 1 for more
detail). The results presented
are found within different experiments:
\begin{itemize}
\item the {$LS$} result is an average value of the best {\em direct} measurement
of the {\sl Lamb splitting\/} ${\cal S}$ by {\em Lundeen and Pipkin\/} \cite{LP} 
and older works.
We also take into account a recent paper by 
{\em van Wijngaaden, Holuj and Drake\/} 
\cite{Wij}, published after our work on review \cite{review} was completed;
\item the {$LS/\Gamma$} result of an indirect measurement of 
the splitting ${\cal S}$.
The measurement was performed by {\em Sokolov and Yakovlev\/} \cite{SY}
for the ratio of  ${\cal S}$ 
and the $2p_{1/2}$
radiative {\em width\/}. The {\em width\/} was recalculated by 
{\sl Karshenboim} 
recently including a leading
radiative correction of relative order $\alpha(Z\alpha)^2\ln(Z\alpha)$
(see Ref. \cite{Comment} for detail)
\[
\Gamma(2p_{1/2})=\frac{2^{10}\pi}{3^8}\,\alpha^3\,Ry\,\frac{m_R}{m}
\left\{1+\ln{\left(\frac{9}{8}\right)}\,\big(Z\alpha\big)^2
+\frac{16}{3\pi}\,\alpha\big(Z\alpha\big)^2\,\big(1.5084...\big)^2\,
\ln{\frac{1}{\big(Z\alpha\big)^2}}
\right\}
\,;
\]
\item in case of {$FS$} the measured value is the {\em fine structure\/} 
splitting 
$2s_{1/2}-2p_{3/2}$, the most recent result for which was found by 
{\em Hagley and Pipkin\/}
\cite{HP};  
\item the {$OBF$} abbreviation means {\em optical beat frequency\/}. 
The value is
an average one from recent {\em Garching\/} \cite{Garching}, 
{\em Yale\/} \cite{Yale} and {\em Paris\/} \cite{Paris} data;
\item the {$CAF$} result is obtained by {\em comparison\/} of 
two {\em optical frequencies\/} measured separately, namely the 
$1s-2s$ interval from
{\em Garching\/} \cite{Udem} and $2s-8s/d$ from {\em Paris\/} 
\cite{P1993,P1997}.
\end{itemize}

\begin{table}[h]
\begin{center}
\begin{tabular}{||c|c||}
\hline\hline &\\[-1ex]
Method & Lamb splitting\\[1ex]
\hline &\\[-1ex]
$LS$ & 1057850(7) kHz  \\[1ex]
$LS/\Gamma$& 1057858(2) kHz\\[1ex]
$FS$ & 1057840(11) kHz \\[1ex]
$OBF$ & 1057843(7) kHz \\[1ex]
$CAF$& 1057853(4) kHz\\[1ex]
\hline\hline
\end{tabular}
\end{center}
\caption{\em Experimental result for the Lamb splitting 
${\cal S}=E(2s_{1/2}-2p_{1/2})$ 
\label{t2}}
\end{table}

\begin{figure}[h]
\epsfxsize=14cm
\centerline{\epsfbox{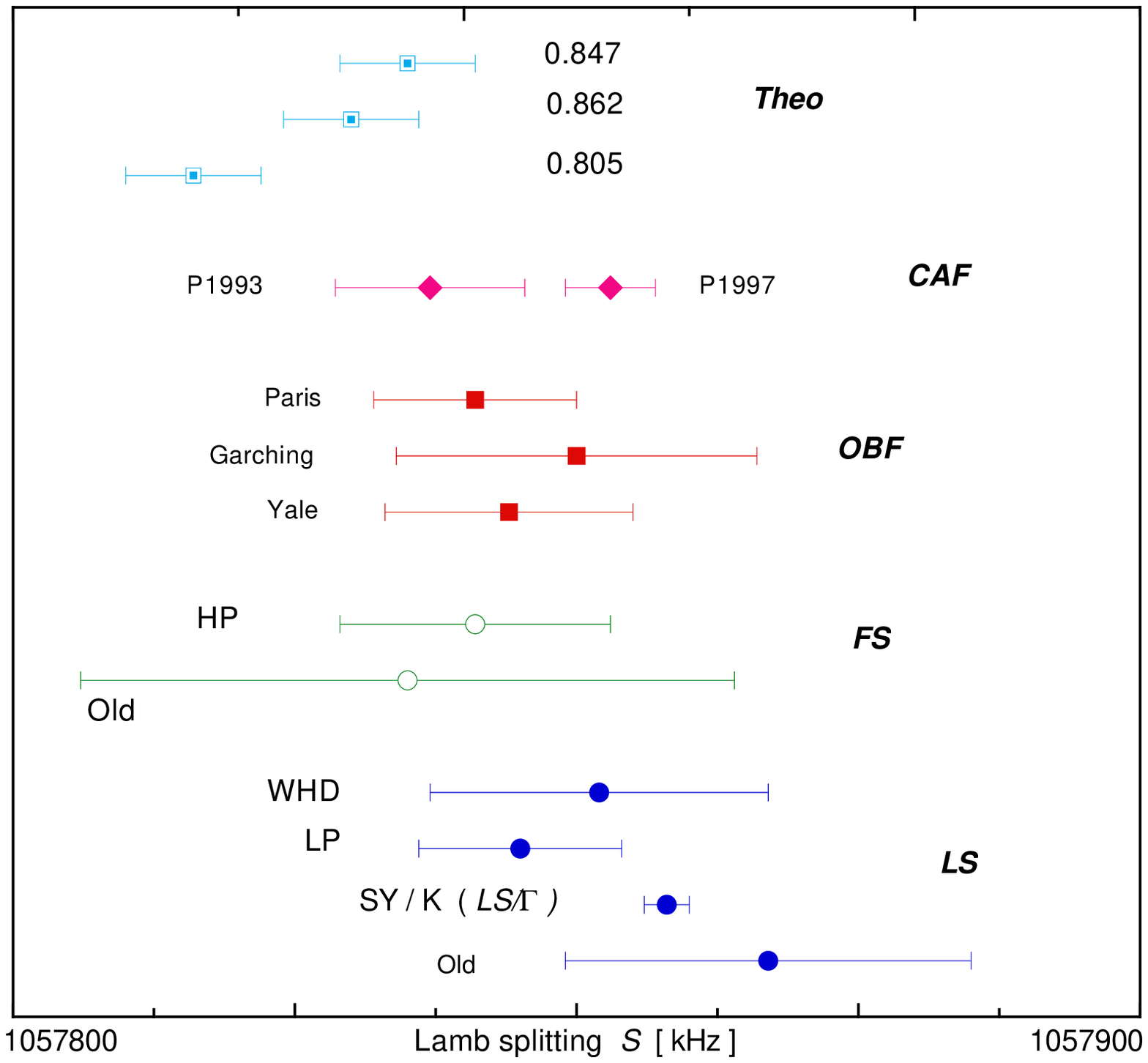}}
\caption{\label{fig1label} Different values of the Lamb splitting ${\cal S}$}
\end{figure}

To recalculate the data obtained by {\em FS\/}, {\em OBF\/} and {\em CAF\/} methods
one has to use a significant piece of theory. In case of the 
{\em fine structure\/} that is the theory of the $2p$ states and in case of the 
{\em optical\/} measurement that is theory of a {\em specific difference\/}
\beq \label{Delta}
\Delta(n)=E_L(1s)-n^3\,E_L(ns)\,.
\eeq
First we explain the importance of this difference and next we consider the
status of its calculation as well as of $E_L(2p_j)$. The progress in measurement
of the either the {\em Lamb splitting\/} or the {\em fine structure\/}
has been relatively
slow. In contrast to that great development was obtained in optical measurement
of the transition frequency between levels with different value of the principal
quantum number $n$ ({\em gross structure\/}). 
The highest precision was reached in two-photon Doppler-free
transitions like $1s\to 2s$. The problem of utilizing those results was due
to the {\em Rydberg constant\/} determination. There is no way to find it except by
investigation of the {\em gross structure\/}. Hence to find anything for 
the Lamb shift one has to measure {\em two\/} optical transitions and 
to construct some difference in which the  {\em Rydberg contribution\/} is
canceled. It is possible
to {\em instrumentally\/} extract a {\em beat frequency\/} directly or
indirectly within an experiment. It is also possible to do that with data obtained
in two {\em independent\/} determinations of different intervals. But still
one has to solve one more problem: the {\em Lamb shift result\/} is after all
a combination of the {\em Lamb shift\/} of the $1s$ and $2s$ states and
also a portion of a higher excited levels contribution is included. 
In order to manage that
a {\em specific difference\/} $\Delta(n)$ was introduced by us some time ago 
(see Ref. \cite{ZP} for detail). This
difference as well as the $2p$ state Lamb shift has a much better theoretical
status than the {\em ground state Lamb shift\/} (or ${\cal S}$). A number
of contributions like e. g. a three-loop term which have not been known up to
date for ${\cal S}$ are known for the difference and for the $2p$ states.

The theoretical expression for the {\em difference\/} of \eq{Delta} is of the form
\beq   \label{LS}
\Delta(n) = \frac{\alpha(Z\alpha)^4}{\pi}\frac{m_R^3}{m^2}\times
\Bigg\{- \frac{4}{3}\ln{\frac{k_0(1s)}{k_0(ns)}}
\left(1+Z\frac{m}{M}\right)^2
\]
\[
+(Z\alpha)^2\times
\Bigg[\left(4\big(\ln(n)-\psi(n+1)+\psi(2)\big)-
\frac{77(n^2-1)}{45 n^2}\right)\ln{\frac{1}{(Z\alpha)^2}}+
A^{VP}_{60}(n)+G^{SE}_{n}(Z\alpha)\Bigg]
\]
\[
-\frac{14}{3}\frac{Zm}{M} \left(\psi(n+1)-\psi(2)
-\ln(n)+\frac{n-1}{2n}\right)\Bigg\}
%
+\frac{\alpha^2 (Z\alpha)^6m}{\pi^2}\ln^2{\frac{1}{(Z\alpha)^2}}\,B_{62}\,.
\eeq
The results of the $2p$ states are similar to that 
\[
\Delta E_L(2p_{1/2})= \frac{\alpha(Z\alpha)^4}{8\pi}\,m\,
\left(\frac{m_R}{m}\right)^3\,\left\{-\frac{4}{3}\ln{k_0(2p)}
\times\left(1+2Z\frac{m}{M}+Z^2(\frac{m}{M})^2\right)\right.
\]
\[
\left.
+(Z\alpha)^2
\left(\frac{103}{180}\ln{\frac{1}{(Z\alpha)^2}}
-\frac{9}{140}+G_{2p_{1/2}}(Z\alpha)\right)
\right\}
- \frac{(Z\alpha)^4\,m}{24}\,m\,
\left(\frac{m_R}{m}\right)^2 )\,
\frac{g_e-2}{2} 
\]
\[
+\frac{\alpha^2 (Z\alpha)^6m}{8\pi^2}\ln^2{\frac{1}{(Z\alpha)^2}}\,B_{62}
+\frac{(Z\alpha)^5}{8\pi}\,m\,\left(\frac{m_R}{m}\right)^3\,\frac{m}{M}
\left(\frac{-7}{18}+(Z\alpha)\frac{\pi}{3}\right)
+\frac{(Z\alpha)^4}{48}\,m\, \left(\frac{m}{M}\right)^2\,
\]
and
\[
\Delta E_L(2p_{3/2})= \frac{\alpha(Z\alpha)^4}{8\pi}\,m\,
\left(\frac{m_R}{m}\right)^3\,\left\{-\frac{4}{3}\ln{k_0(2p)}
\times\left(1+2Z\frac{m}{M}+Z^2(\frac{m}{M})^2\right)\right.
\]
\[
\left.
+(Z\alpha)^2
\left(\frac{29}{90}\ln{\frac{1}{(Z\alpha)^2}}
-\frac{1}{70}+G_{2p_{3/2}}(Z\alpha)\right)
\right\}
+ \frac{(Z\alpha)^4\,m}{48}\,m\,
\left(\frac{m_R}{m}\right)^2 )\,
\frac{g_e-2}{2} 
\]
\[
+\frac{\alpha^2 (Z\alpha)^6m}{8\pi^2}\ln^2{\frac{1}{(Z\alpha)^2}}\,B_{62}
+\frac{(Z\alpha)^5}{8\pi}\,m\,\left(\frac{m_R}{m}\right)^3\,\frac{m}{M}
\left(\frac{-7}{18}+(Z\alpha)\frac{\pi}{3}\right)
-\frac{(Z\alpha)^4}{96}\,m\, \left(\frac{m}{M}\right)^2\,.
\]
The one-loop contribution
due to the {\em vacuum polarization\/} ($A^{(VP)}_{60}$) was found for an 
arbitrary 
$ns$ state by {\em Karshenboim and Ivanov\/} and for an $np_j$ state by {\em
Manakov, Nekipelov and Feinstein\/}. The {\em self-energy\/} term 
($G^{SE}_{n}(Z\alpha)$) is mainly determined by a value of $A^{(SE)}_{60}$. 
The last
was found by 
{\em Pachucki\/} for $1s$ and $2s$ and by {\em Jentschura and Pachucki\/} 
for $2p$. The result has to be found extrapolating numerical data calculated
by {\em Mohr\/} and by {\em Kim and Mohr\/} 
for higher nuclear charge $Z$. It is simpler
to extrapolate the data for the {\em difference\/} of \eq{Delta} and for the 
$p$ state 
in comparison to extrapolation for the ground state.
The {\em leading two-loop logarithmic term\/} was found by {\em Karshenboim\/}.
{\em Pachucki and Grotch\/} found that the {\em recoil\/} term of the order
$(Z\alpha)^6m^2/M$ for an $ns$ state is scaling by $1/n^3$ and so it 
cannot contribute
to the {\em difference\/}. In case of the $2p$ states this correction 
was evaluated by
{\em Golosov, Yelkhovsky, Milstein and Khriplovich\/}.

The theoretical progress mentioned for the $\Delta(n)$ and the $2p$ Lamb shift 
allows to obtain results in Table 2 with pretty small {\em theoretical 
uncertainty\/}. 
The QED results for $4p$ states are also needed for the evaluation of 
some experimental data. 
The value of $A^{(SE)}_{60}$ was found by {\em Jentschura, Mohr and Soff\/}
and the {\em recoil\/} effects were investigated by {\em Yelkhovsky\/} and 
{\em Pachucki\/} independently.
That is, however, only a {\em not} significant part of the 
QED calculations which have to be done
on a way to reach a value of the
proton radius from hydrogen atom spectroscopy. It is necessary 
to calculate the Lamb shift
of the $2s$ state (as we can see from the discussion above the 
significant part of the QED theory
of $1s$ and $2s$ states is the same and we speak here about the ground state).

The expression for the {\em ground state Lamb shift\/} 
is much more complicated. The term most important for the further 
discussion is of the well known form
\beq \label{radius}
\delta E_{nucl}(ns) = \frac{2}{3}\frac{(Z\alpha)^4m_R}{n^3}
m_R^2 \, R_p^2\,.
\eeq
That includes the proton charge radius and subtracting the QED result 
(\eq{radius} excl.) from
the experimental value we can determine the proton charge radius.
A detailed review of the ground state
Lamb shift was presented by {\em Pachucki et al.\/} \cite{Pachuckietal}. 
Here we mention some important features. 
The most recent corrections
considered there are the {\em one\/}-loop {\em self-energy\/} ({\sl Pachucki\/};
{\em Mohr\/}), 
the {\em two\/}-loop 
contribution of the order $\alpha^2(Z\alpha)^5m$ 
({\em Pachucki\/} and {\em Eides, Grotch and Shelyuto\/}), 
the {\em leading\/} two-loop {\em logarithm\/} of 
a {\em higher\/} order ({\em Karshenboim\/}), 
a {\em pure recoil\/} term ({\em Fell et al.\/};
{\em Grotch and Pachucki\/} and {\em Shabaev and coworkers\/}) 
and the {\em radiative-recoil\/} correction ({\em Bhatt and 
Grotch\/} contradicting to {\em Pachucki\/}).

We mainly agree with a consideration in review \cite{Pachuckietal}. However, 
our result is {\em shifted\/} by -3.6 kHz because the
$\alpha^2(Z\alpha)^6 m\log^3{Z\alpha}$-correction 
has {\em not} been included there. We would also like to present here
all sources of the theoretical uncertainty for the Lamb splitting 
${\cal S}$:
\begin{itemize}
\item[$\bullet$] unknown $\alpha(Z\alpha)^7 m$ and higher order {\em one-loop} corrections
are estimated to 1 kHz;
\item[$\bullet$] $\alpha^2(Z\alpha)^6 m \log^2{Z\alpha}$ and higher order
{\em two-loop\/} terms can give up to 2 kHz;
\item[$\bullet$] the {\em three-loop} $\alpha^3(Z\alpha)^5 m$ contribution is here
estimated preliminary to 2
kHz and needs more understanding.
\end{itemize}
The one-loop term is going to be calculated exactly and so the uncertainty is
being removed \cite{PC}. 

\begin{figure}[h]
\epsfxsize=14cm
\centerline{\epsfbox{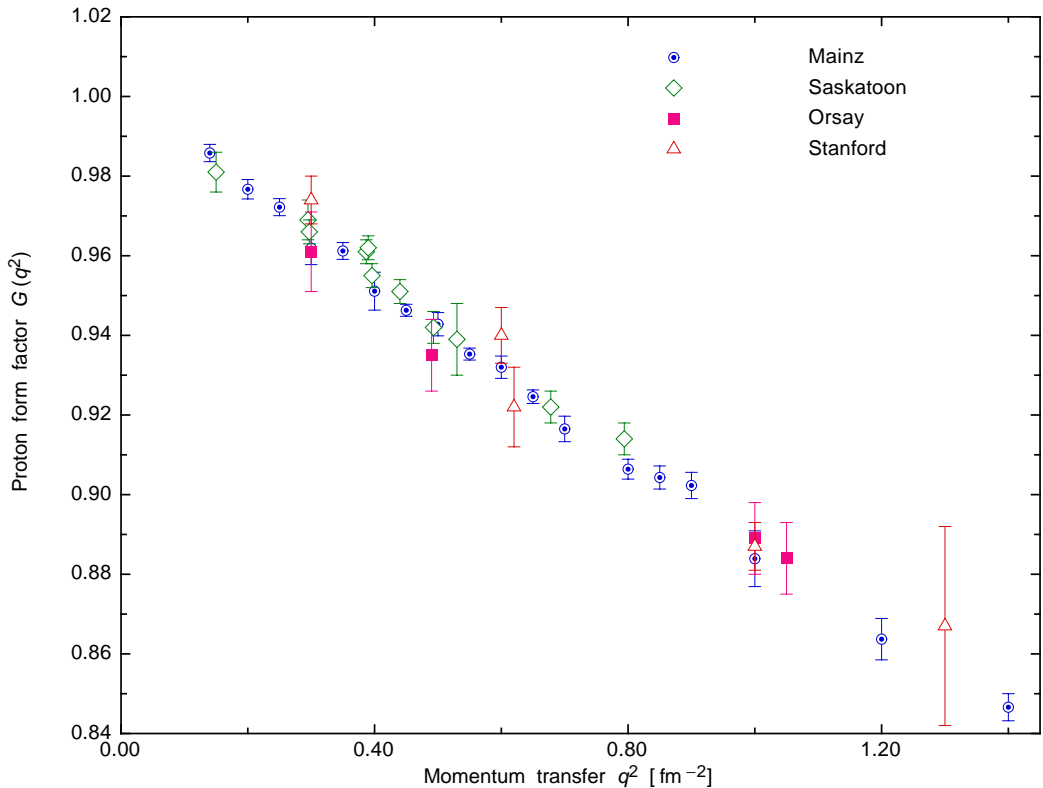}}
\caption{\label{fig2label} Experimental data for the proton electric form factor}
\end{figure}

Now let us turn to the discussion of the {\em scattering\/} 
results. We start from simple estimates.
The value {\em extracted\/} from {\em elastic electron-proton\/} 
scattering cross sections is the {\em electric form factor\/} 
of the proton $G(q^2)$. We need to 
investigate it at {\em low momentum transfer\/} 
where the {\em `signal'} which is $G-1$ is mainly 
determined by the radius term ($(q^2R^2_p)/6$). The $G-1$ term lies 
at the condition of 
{\em experiment\/} in Ref. \cite{Simon} between 1\% and 15\% (see Fig. 2). 
The {\em scattering\/} radius in Table 1 is claimed \cite{Simon} 
to be with uncertainty 
within about 1\%. That means that
all shifts on the level of a few percents are important. 
We discussed in our review \cite{review}
some {\em QED\/} corrections to the {\em cross section\/} 
which are expected to be on a 
few percent level and which are beyond
evaluation of Ref. \cite{Simon}. In this 
paper we concentrate our attention to more 
important problem. That is due to the {\em normalization\/} of the data. 
It is clear that the value of
the electric form factor at zero momentum transfer is equal to one 
\[
G_{th}(0)=1\,.
\]
That is absolutely correct {\em theoretically\/}, but the form 
factor {\em cannot\/} be measured 
{\em straightforwardly\/}. It is {\em possible\/} 
to measure only some cross section. Evaluating the data
one can extract form factor from the Rosenbluth formulae. 
On this way the {\em  extracted\/} experimental
value which is expected to be the {\em form factor\/}
is only {\em consistent\/} with the true form factor within some 
{\em uncertainty\/}. As one of
results the value of $G_{exp}(0)$ is consistent with one and it has to be 
found within
some fitting procedure. In other words we have to write a trivial equation
\[
G_{exp}(q^2)=a_0\, G_{th}(q^2)\,,
\]
where the value of $a_0$ is consistent with one, 
but not equal to that {\em a priori\/}. 
For application to the low momentum transfer electron-proton scattering
\cite{Simon} we can
expect that $a_0$ is a constant (but it realy 
{\em depends\/} on experimental setup).

\begin{figure}[h]
\epsfxsize=14cm
\centerline{\epsfbox{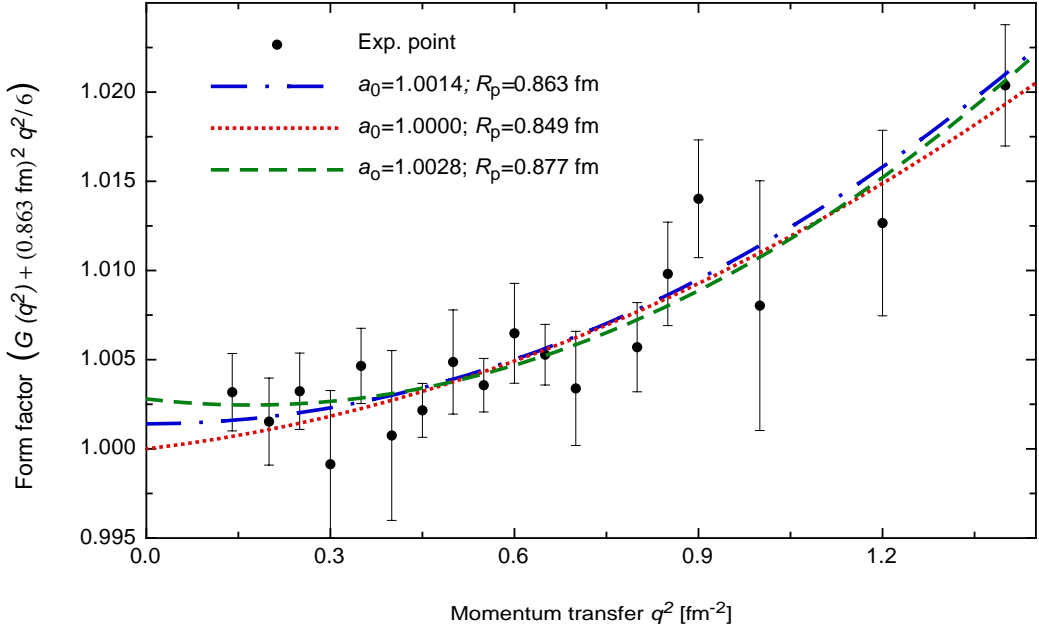}}
\caption{\label{fig3label} Some fitting of the {\em Mainz\/} data with 
different normalization constant}
\end{figure}

In both {\em Mainz\/} papers \cite{Simon,Mergell} some special {\em a priori\/} 
prescriptions for $a_0$ were used.
The problem of {\em normalization\/} was investigated by {\em Wong\/} 
\cite{Wong} prior to
publication of the dispersion paper \cite{Mergell}. 
He found that the result {\em strongly depends\/} on an assumption
on the value of the constant $a_0$. The correct value of the radius 
corresponds to the {\em free normalization\/} 
(i. e. $a_0$ is one of the fitting parameters). The constant
\[
a_0=1.0028(22)
\]
found from the Mainz scattering data \cite{Simon} {\em a posteriori\/} \cite{Wong}
(see Fig. 3)
is consistent with one (the prescription of Ref. \cite{Mergell} is $a_0=1$) and
with the value of $a_0=1.0014$ prescripted in Ref. \cite{Simon}. 
However, on this way the uncertainty of 
the proton radius is significantly larger than in both Mainz papers \cite{Simon,Mergell}
(see Ref. \cite{review} for more details).

We have discussed all items of Table 1 and we present now some {\bf conclusions}. 
The scattering 
value has an uncertainty of about 0.24 fm. We {\em cannot\/} 
present here any {\em eventual\/} figure because
{\em Wong\/} performed his evaluation with 
the most important part of data but not with all of them. 
Not all corrections to the data were included. The result should be close to the Wong value
(0.877(24) fm) but it needs more analysis. However data are quite old and practically it
is not possible to reevaluate them. Hopefully, a 
{\em new} measurement in Mainz is going to be done.
The Lamb shift examination {\em can\/} give a {\em better\/} 
value. We have to mention that the experiment
{\em claimed\/} as the {\em most accurate\/} one
(uncertainty is 2 kHz) is under some criticism and the average value
of all other measurements gives now some larger uncertainty (about 3 kHz).
The uncertainty for the proton radius extracted is 0.15 fm (the equal portions arise from the
ground state Lamb shift {\em QED theory\/} 
and from the {\em optical experiments\/}). We also would {\em not\/} like to
give any value because it is a running value now: some data are still coming. The most precise
value which is included into our analysis is due to the {\em Garching\/} 
and {\em Paris\/} absolute measurements. Some parts of the {\em Paris\/} data
are changing now due to recalibration of the standards \cite{PC1}. The preliminary value from
the Lamb shift is in {\em fair agreement\/} with the Wong value but with 
{\em twice\/} larger accuracy.

The situation now, when the uncertainties from 
{\em spectroscopy\/} measurement, {\em QED\/} calculation and
the {\em scattering\/} data are on the same level 
{\em is\/} quite challenging and we believe that new results
are coming in all these fields. 
Recently a new wave of activity
projecting of a {\em  muonic hydrogen\/} experiment 
for the Lamb shift has been arisen at the {\em PSI\/} and we also hope     
that will give one more value for the proton radius.
We also would like to attract attention to the Lamb shift measurement
in hydrogen-like system with a {\em moderate value\/} 
of the nuclear charge $Z\simeq 5-20$.
It seems it is necessary to encourage such investigations because
they can lead to {\em experimental\/} estimation of the {\em higher order
QED\/} terms. That is to be helpful for the {\em hydrogen\/} Lamb shift theory.\\

\medskip

This work was supported in part by the Russian state program of 
{\em Fundamental metrology}. We would like to thank T. W. H\"ansch 
and the whole hydrogen team of the
MPQ for the support, hospitality and stimulating discussions.


\vspace*{.5cm}
\begin{thebibliography}\\
\bibitem{Fell}  R. N. Fell, I. B. Khriplovich, A. I. Milstein and A. S.
Yelkhovsky, {\it Phys.  Lett.\/} A{\bf 181} (1993) 172.
\bibitem{Pachu} K. Pachucki, {\it Phys. Rev.\/} A{\bf 56} (1997) 297; 
{\it Phys. Rev. Lett.\/} {\bf 79} (1997) 4120.
\bibitem{review} S. G. Karshenboim, What do we actually know on the proton radius?
{\em MPQ report\/} 230 (1998), e-print {\em hep-ph\/}/9712347, to be published
in {\it Can. J. Phys\/}.
\bibitem{Hand}
L. Hand, D. I. Miller and R. Willson, {\it Rev. Mod. Phys.\/}
{\bf 35} (1963) 335.
\bibitem{Simon}
G. G. Simon,  Ch. Schmitt F. Borkowski and V. H. Walther,
{\it Nucl.  Phys.\/}  A{\bf 333} (1980) 381.
\bibitem{Draper}
T. Draper, R. M. Woloshin, and K.-F. Liu, {\it Phys. Lett.\/} {\bf 234}B (1990) 121.
\bibitem{Leinweber93}
D. B. Leinweber and T. D. Cohen, {\it Phys. Rev.\/} D{\bf 47} (1993) 2147.
\bibitem{Mergell}
P. Mergell, U.G. Meissner and D. Drechsel, {\it Nucl. Phys.\/} A{\bf 596} (1996) 367.
\bibitem{Udem}
T. Udem, A. Huber, B. Gross, J. Reichert, M. Prevedelli,
M. Weitz and T. W. H\"ansch, {\it Phys. Rev. Lett.\/} {\bf 79} (1997) 2646.
\bibitem{LP} S.~R.~Lundeen and F.~M.~Pipkin, {\it Phys. Rev. Lett.\/}  {\bf
46} (1981) 232; {\it Metrologia\/} {\bf 22} (1986) 9.
\bibitem{Wij} A. van Wijngaaden, F. Holuj and G. W. F. Drake, {\it Can. J. Phys.\/}
{\bf 76} (1998) 95.
\bibitem{SY} Yu.~L.~Sokolov and V.~P.~Yakovlev, {\it Sov. Phys. JETP\/} {\bf
56} (1982) 7.
\bibitem{Comment} S. G. Karshenboim, {\it Phys. Sc.\/} {\bf 57} (1998) 213.
\bibitem{HP} E. W. Hagley and F. M. Pipkin, Phys. Rev. Lett. {\bf 72},
1172 (1994).
\bibitem{Garching} M. Weitz, A. Huber,  F. Schmidt-Kaler,
D. Leibfried and T. W. H\"ansch, {\it Phys.  Rev.  Lett.\/}  {\bf 72} (1994) 328.
\bibitem{Yale} D. J. Berkeland, E. A. Hinds and M. G. Boshier,
{\it Phys. Rev. Lett.\/} {\bf 75} (1995) 2470.
\bibitem{Paris} S. Bourzeix, B. de Beauvoir, F. Nez, M. D. Plimmer,
F. de Tomasi, L. Julien, F. Biraben and D. N. Stacey,
{\it Phys. Rev. Lett.\/} {\bf 76} (1996) 384.
\bibitem{P1993} F. Nez, M. D. Plimmer, S. Bourzeix, L. Julien, F. Biraben,
R. Felder, Y.  Millerioux and P. De Natale, {\it Europhys.  Lett.\/} {\bf 24}
(1993) 635.
\bibitem{P1997} R. de Beauvoir, F. Nez, B. cagnac, F. Biraben,  D.
Touahri, L. Hilico, O. Acef, A. Clairon and J. J. Zondy,
{\it Phys. Rev. Lett.} {\bf 78} (1997)
440.
\bibitem{ZP} S. G. Karshenboim, {\it Zeit. Phys.} D{\bf 39} (1997) 109.
\bibitem{Pachuckietal} 
K. Pachucki, D. Leibfried, M. Weitz, A. Huber, W.
K\"onig and T. W. H\"ansch, {\it J. Phys.\/} B{\bf 29} (1996) 177; {\sl Corrigendum},
{\sl ibid}., 1573.
\bibitem{PC} U. Jentschura, P. Mohr and G. Soff, {\em private communication}.
\bibitem{Wong} Ch. W. Wong, {\it Int. J. Mod. Phys.\/} {\bf 3} (1994) 821.
\bibitem{PC1} F. Biraben and F. Nez, {\em private communication}.
\end{thebibliography}
\end{document}